# Rydberg-State Hopping in a Wavemeter-Locked Dissipative Time-Crystal System

D. Arumugam

*Jet Propulsion Laboratory, California Institute of Technology*

(Electronic mail: darmindra.d.arumugam@jpl.nasa.gov.)

(Dated: 6 November 2025)

Rydberg-state hopping is demonstrated in a wavemeter-locked two-photon rubidium system (Rb $D_2$ probe at 780 nm and 480 nm coupler), enabling rapid and repeatable switching between the $65S_{1/2}$ and $63D_{5/2}$ states without cavity or frequency-comb stabilization. A Fizeau-interferometer wavemeter provides the error signal for a digital feedback loop that simultaneously stabilizes the coupler and commands discrete Rydberg-state changes. The lock achieves sub-MHz frequency stability and acquisition rates up to 6.5 GHz s$^{-1}$ (0.4283 GHz engaged in 66 ms), extrapolating to ~0.93 s for a ~6 GHz $65S\leftrightarrow63D$ transition. Time-resolved spectra reveal re-emergent dissipative time-crystal oscillations after each hop, with distinct state-dependent fundamentals and harmonics. This approach addresses the need for dynamically reconfigurable Rydberg-state control for on-resonant multi-band field detection, while the DTC frequency reconfigurability enables adaptive, low-frequency E-field sensing in compact, cavity-free architectures.

Electrometry using Rydberg atoms enables broadband, SI-traceable measurements from RF-to-THz by exploiting giant dipole moments and ladder transitions of high-n states [1–4]. Despite progress, a system limitation persists: the inability to dynamically tune between discrete Rydberg states while continuously frequency locked. Existing approaches to address this fall into two categories: Multi-tone RF that access multiple transitions simultaneously or sequentially [5, 6], but splits atom population among states and requires wideband (>100 GHz) RF sources, limiting practicality; Frequency-comb lasers that generate multiple carriers to address different transitions [7, 8], achieving sub-100 μs switching [9] but relying on complex comb-generation—high-frequency electro-optic modulator (EOM) chains, filtering, and phase-locked stabilization, limiting deployability. Direct EOM/AOM (acousto-) frequency translation is challenging—AOMs offer < 1 GHz shift and single-sideband EOMs are difficult to realize at 480/510 nm. To date, no compact, single loop actively stabilized laser system has demonstrated dynamic tuning between Rydberg-states. The approach here employs a Fizeau-interferometer wavemeter [10, 11] as both frequency reference and control element to stabilize and enable high-speed digital set-point control to achieve state hopping without combs, cavities, or auxiliary phase locks — offering a simple route to on-resonant Rydberg state tuning.

Two-photon Rydberg ladder systems (probe and coupler) underpin electrometry with Rydberg atoms [12, 13]. The probe is frequently stabilized to hyperfine structure through saturated-absorption spectroscopy (SAS) [14]. Frequency stabilization of the coupler is through electromagnetically induced transparency (EIT) locking [15], or Pound–Drever–Hall (PDH) cavity techniques [16, 17]. While these deliver sub-MHz-to-kHz stabilization, they constrain rapid tuning and increase optical, mechanical, and electronic overhead. In contrast, commercial Fizeau-based wavemeters can provide absolute wavelength readout with a resolution <1 MHz, 1 kHz rates, and no moving parts [18]. Integrated into a digital feedback loop [10], the wavemeter output can drive coupler-laser PZT (piezo transducer), enabling sub-MHz stability; Fizeau-wavemeter locks have demonstrated Allan deviation <10$^{-9}$ at $10^3$ s integration [10]. In the power-broadened regime of ladder-EIT, the effective two-photon linewidth is $\Gamma_{2ph}\sim\Omega_c^2/\Gamma_e$ which for $\Omega_c/2\pi > 10$ MHz ( $\Gamma_e/2\pi \sim 6$ MHz) yields $\Gamma_{2ph} > 16$ MHz; Keeping coupler frequency noise to sub-MHz ensures that laser fluctuations remain negligible compared with the two-photon linewidth. Here, a Fizeau-wavemeter lock is used to execute repeatable hops between Rydberg states ($65S_{1/2}$ and $63D_{5/2}$)— with acquisition rates up to 6.5 GHz s$^{-1}$ and lock within 1s, establishing a simple single loop digitally reconfigurable Rydberg-state system.

Recent advances in dissipative time-crystal (DTC) dynamics in Rydberg ensembles reveal a nonequilibrium regime of self-sustained oscillations (OSC) arising from many-body interactions across the Rydberg manifold [19–21]. These oscillations exhibit Rydberg-state dependent fundamental frequencies [19] and have achieved state-of-the-art E-field sensitivities in the kHz and sub-kHz bands tied to (injection locked) the OSC frequency [22, 23]. Integrating DTC operation with state hopping offers two capabilities: (1) tunable selection among Rydberg-states whose oscillation (OSC) frequencies correspond to different sensing frequencies, and (2) a direct probe of lock-acquisition dynamics, as the DTC signatures appear only after full stabilization. The demonstrated wavemeter-locked two-photon Rb $D_2$ system (780/480 nm) achieves stable Rydberg-state hopping with re-emergent DTC oscillations following each hop, establishing a compact route toward Rydberg-state–reconfigurability.

I. RYDBERG-STATE HOPPING EXPERIMENT

A two-photon Rydberg ladder system is configured using the $5S_{1/2}\to 5P_{3/2}\to nS/nD$ transitions of $^{87}$Rb (Fig. 1a,b). A weak probe beam at 780 nm and a strong coupler beam at 480 nm counter-propagate through a room-temperature vapor cell (length =56 mm, diameter =25 mm). The 480 nm light is produced by frequency-doubling a tunable 960 nm

# Rydberg-State Hopping in a Wavemeter-Locked Dissipative Time-Crystal System

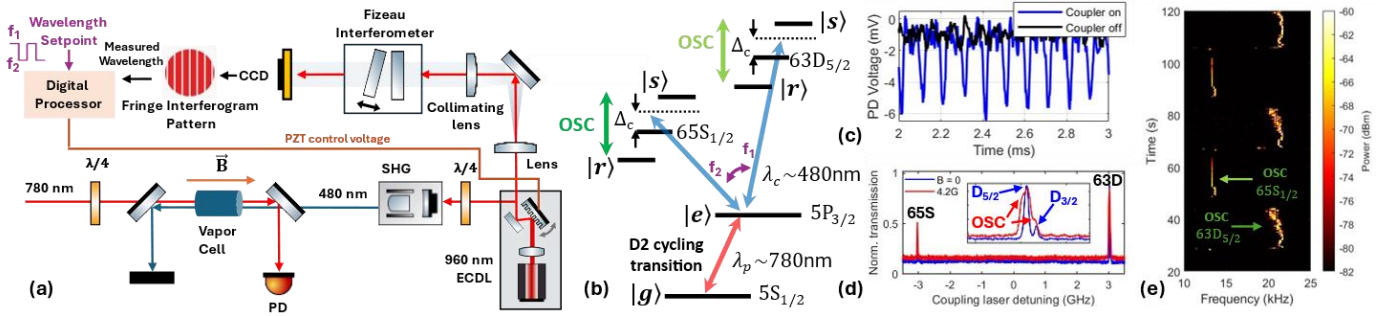

**Figure 1: Two-photon Rydberg-state hopping experiment.** (a) Setup: a 960 nm coupler master laser is stabilized and rapidly tuned between two Rydberg resonances (65S ↔ 63D) using a high-speed wavelength-meter feedback loop. A digital processor generates setpoints ($f_1$, $f_2$) and drives the laser PZT through a custom PI controller; the same wavelength-meter signal provides both lock and tuning, removing the need for a cavity or PDH system. The 780 nm probe counter-propagates with the 480 nm coupler (frequency-doubled 960 nm) through a Rb vapor cell in a controlled magnetic field, and probe transmission is detected on a photodiode. (b) Simplified energy-level diagram showing the two-photon ladder excitation $5S_{1/2} \rightarrow 5P_{3/2} \rightarrow nS/nD$, where the coupler alternately addresses $65S_{1/2}$ and $63D_{5/2}$ states to realize Rydberg-state hopping. (c) Time-domain probe signal showing oscillatory dynamics of the $65S_{1/2}$ state, characteristic of a Rydberg dissipative-time-crystal (DTC) regime (blue trace is coupler on, black with coupler off). (d) Normalized transmission versus coupler detuning over ±3.5 GHz, revealing $63D_{3/2}$, $63D_{5/2}$, and $65S_{1/2}$ EIT resonances. The red trace (B = 4.2 G) exhibits oscillatory signals (OSC) at both D states, as shown in the inset. (e) Spectrogram of the probe response showing discrete oscillation bands corresponding to $65S_{1/2}$ and $63D_{5/2}$ manifolds, demonstrating wavelength-meter-controlled Rydberg-state hopping.

diode laser in a resonant cavity containing a PPKTP (quasi-phase-matched KTP) crystal. The probe is frequency-stabilized to the $5S_{1/2}$, F = 2→$5P_{3/2}$, F' = 3 $|g\rangle \rightarrow |e\rangle$ cyclic transition via Doppler-free pump-probe SAS, ensuring < 100 kHz stability. Both beams have ≈1 mm 1/e² diameters, optical powers of 162 µW (probe) and 438 mW (coupler), and free-running instantaneous linewidths of < 90 kHz. A uniform bias magnetic field (4.2 G via a Helmholtz coil) applied along the beam axis lifts Zeeman degeneracy and defines a quantization direction, enabling selective coupling between magnetic sublevels that support DTC oscillations [19, 22]. DTC dynamics manifest as sustained oscillations between $|r\rangle$ and nearby sublevels $|s\rangle$ (Fig. 1b), governed by Zeeman splitting from the bias field and polarization-dependent selection rules. Quarter-wave plates are used to optimize elliptical components of both fields to enhance DTC OSC amplitude [19]. The dynamic frequency control loop (Fig. 1a) employs a Fizeau-interferometer wavemeter (HighFinesse WS-series) as the optical reference, performing multi-fringe acquisition and frequency/phase extraction (Fig. 2a) in a thermal and pressure-stabilized Fizeau interferometer with fused-silica etalons. The wavemeter measures the 960 nm at 1 kHz rate and < 1 MHz resolution, feeding a digital proportional–integral (PI) controller in software (Fig. 2a). The feedback drives the laser piezoelectric transducer (PZT) directly, maintaining lock and enabling high-speed commanded wavelength steps. Locked to the $65S_{1/2}$, the resulting probe transmission (Fig. 1c) displays DTC oscillations at ≈10 kHz that vanish when the coupler is blocked, verifying optical-field-driven many-body dynamics. Figure 1d shows the probe transmission as the coupler is scanned over 7 GHz, revealing distinct EIT resonances corresponding to the $63D_{3/2}$, $63D_{5/2}$, and $65S_{1/2}$ states; the red trace (B = 4.2 G) exhibits additional oscillatory features

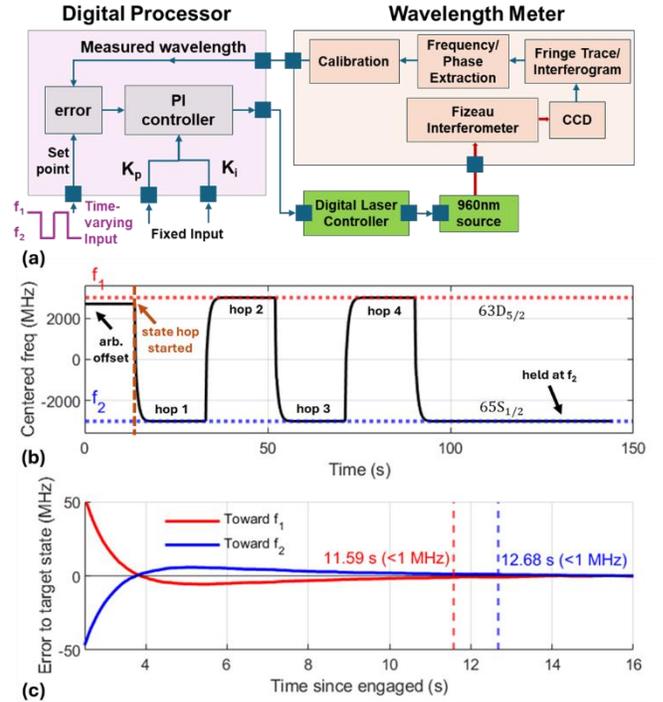

**Figure 2: Wavemeter-locked feedback system for Rydberg-state hopping.** (a) Control architecture for the wavemeter feedback. The measured wavelength is processed through a proportional–integral (PI) controller with $K_I/K_P = 1$, driving the 960 nm coupler laser via a digital controller. The Fizeau interferometer–based wavemeter performs fringe acquisition, phase extraction, and calibration to provide frequency estimates. A time-varying setpoint sequence ($f_1$, $f_2$) defines the hopping targets corresponding to $65S_{1/2}$ and $63D_{5/2}$. (b) Time evolution of the laser frequency showing repeated transitions ("hops") between the two setpoints. The controller maintains lock after the transition, demonstrating reproducible and stable hopping. (c) The system settles to within 1 MHz of the target frequency in 11.6 s for the upward hop (red) and 12.7 s for downward hop (blue), confirming symmetric dynamics.

Rydberg-State Hopping in a Wavemeter-Locked Dissipative Time-Crystal System

arising from Zeeman-split sublevels that sustain DTC oscillations, as highlighted in the inset for 63D. Two digital setpoints ($f_{1,2}$) are used corresponding to the $63D_{5/2}$ and $65S_{1/2}$ manifolds separated by about ~6.007 GHz, where precise wavelengths are determined empirically and coupler detuning found to be $\Delta_c/2\pi \approx +20$ MHz. Sequential switching between $f_1$ and $f_2$ establishes deterministic Rydberg-state hops, with the same wavemeter simultaneously acting as frequency reference and error sensor. Example oscillation spectra (Fig. 1e) show state-dependent fundamentals—around 22 kHz for $63D_{5/2}$ and 10 kHz for $65S_{1/2}$—with reappearance after every hop once lock is re-established.

Figure 2b traces the coupler-laser frequency during repeated state hops (starting at an arbitrary but wavemeter PI stabilized wavelength). Each transition reproduces to within < 1 MHz of the setpoint, confirming closed-loop stability. The up- and down-hops (63D → 65S and back) settle in 11.6 s and 12.7 s, respectively (Fig. 2c), in this case, simply limited by PI controller-gain ($K_I/K_P = 1$) rather than loop bandwidth or piezo response rates. Together, Figs. 1–2 demonstrate a self-contained wavemeter-locked platform capable of stabilizing and digitally reconfiguring Rydberg excitations between distinct manifolds with sub-MHz precision and no cavity or frequency-comb infrastructure.

## II. RESULTS

To examine the emergence of DTC oscillations as the coupler laser was digitally programmed to hop between the $63D_{5/2}$ and $65S_{1/2}$ manifolds, the capture of OSC dynamics was synchronized with wavemeter measurements. Figure 3a shows a time–frequency spectrogram of the probe signal, revealing alternating oscillation bands synchronized with each programmed wavelength hop. These bands correspond to self-sustained DTC oscillations, whereas far-detuned intervals (red) show only detector noise. The oscillation frequencies (≈10 kHz for $65S_{1/2}$ and 22 kHz for $63D_{5/2}$)— reflect the differing many-body coupling strengths of each manifold. Their reappearance with each hop demonstrates a robust, state-dependent time-crystalline order. The synchronized wavemeter trace in Fig. 3b confirms precise, periodic detuning of the coupler between the two manifolds. The timing of these hops aligns exactly with the onset of the oscillation bands in Fig. 3(a). Frequency-domain slices in Fig. 3c highlight this behavior at representative times. Before DTC engagement (t = 10 s, blue), the spectrum is noise-limited; once tuned into resonance (t = 40 s, yellow; t = 55 s, red), pronounced peaks emerge at the fundamental oscillation frequencies of $65S_{1/2}$ and $63D_{5/2}$. The presence of higher-order harmonics confirms nonlinear mode coupling intrinsic to the DTC regime. These harmonics persist across successive state hops without external retriggering, indicating a rapidly stabilized temporal order within the driven ensemble.

The transient response of the wavemeter-locked coupler during activation of DTC oscillations is given in Fig. 4. In Fig. 4a, the probe spectrogram shows the coupler engaging $65S_{1/2}$ from 0.43 GHz detuning, where fundamental and harmonic oscillation bands emerge shortly after each engagement (yellow dashed lines), marking DTC onset and stabilization. The corresponding wavemeter traces in Fig. 4b display sequential lock steps with PI gain ratios $K_I/K_P$=3–30. Gray dashed lines denote PI engagement, while yellow lines indicate delayed DTC onset. Figure 4c overlays the transients, showing rise times decreasing from 1.11 s to 0.066 s with increasing $K_I/K_P$. Higher ratios yields faster stabilization but introduces mild ringing before lock settles. This demonstrates that controller-gain tuning governs both lock response and settling time delays that drive emergence of DTC oscillations.

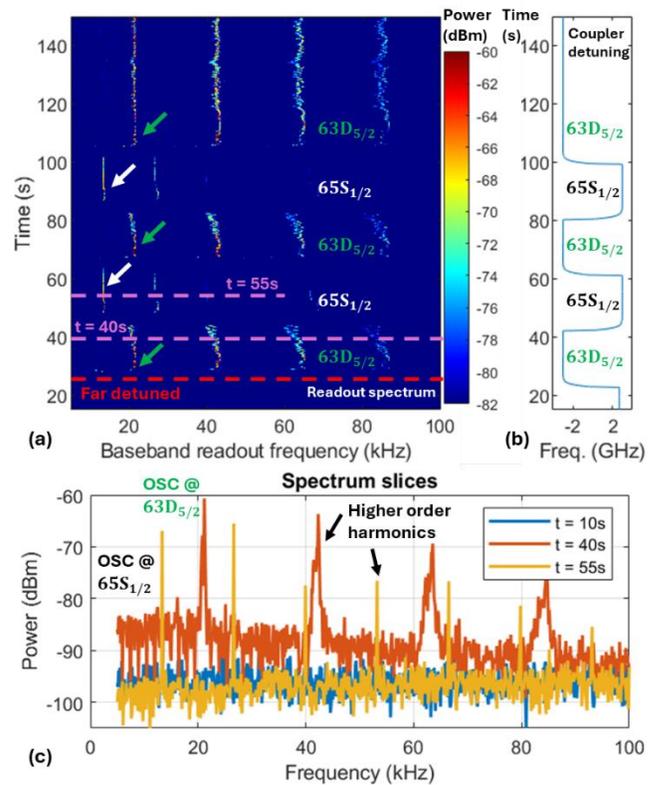

**Figure 3: Dissipative time-crystal (DTC) dynamics during Rydberg-state hopping.** (a) Spectrogram of the probe photodiode signal showing temporal evolution of the DTC oscillation spectrum as the coupler laser hops between Rydberg states. Distinct oscillation bands appear alternately at the $63D_{5/2}$ and $65S_{1/2}$ manifolds (white vs. green arrows), synchronized with the programmed wavelength hops. Each band corresponds to self-sustained oscillations in the DTC regime, while initial intervals labeled far-detuned (red) show only background noise. (b) Simultaneous wavemeter readout of the coupler-laser detuning illustrating controlled wavelength hops between $63D_{5/2}$ and $65S_{1/2}$. The timing of each hop matches the DTC transitions in (a). (c) Frequency-domain slices of the readout at times: t = 10 s (blue) shows noise before DTC engagement; t = 40 s (yellow) and t = 55 s (red) exhibit strong oscillatory peaks at the $63D_{5/2}$ and $65S_{1/2}$ states, respectively, along with higher-order harmonics characteristic of nonlinear DTC dynamics.

Figure 5 shows that the lock acquisition rate for the $65S_{1/2}$ state follows a power-law dependence R=0.08 $(K_I/K_P)^{1.2}$, increasing predictably with controller gain. With an initial detuning of 0.43 GHz, rates up to ≈6.5 GHz s$^{-1}$ are achieved, indicating that the loop is gain-limited rather than measurement-limited. Ringing at high gain remains mild and well-damped and could be further suppressed by adding a derivative term. The ultimate speed is bounded by the 1 kHz wavemeter update rate and PZT actuator bandwidth, allowing fine-lock acquisition on the order of tens of GHz s$^{-1}$ within the actuator's linear range.

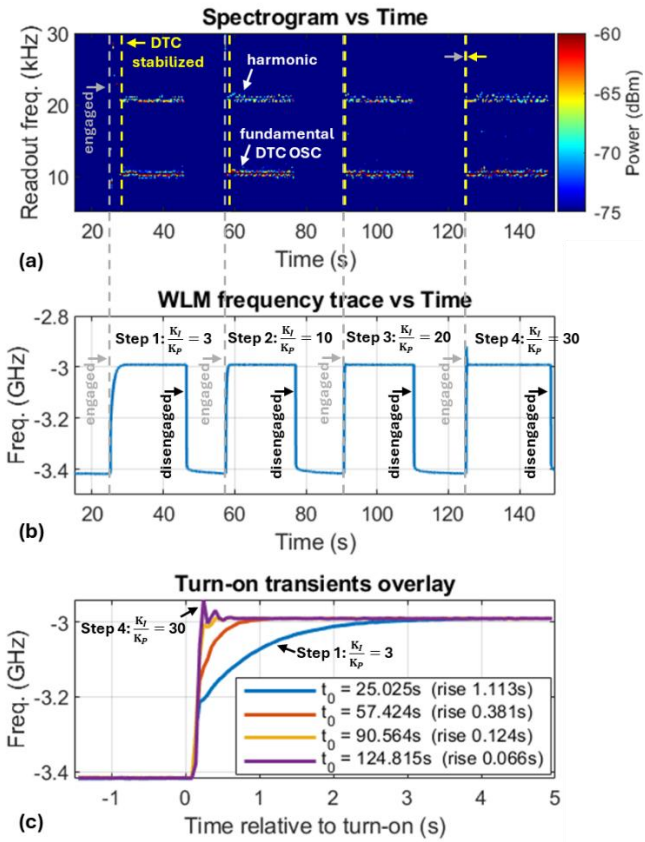

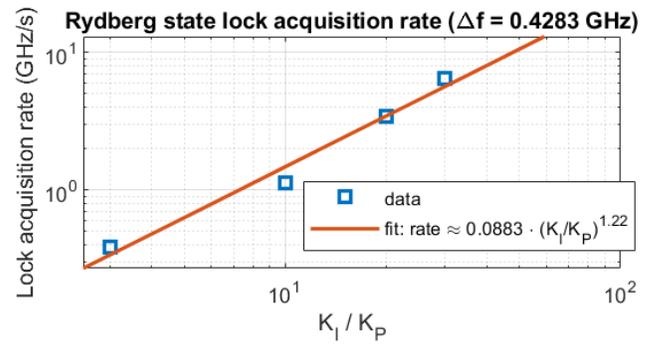

**Figure 4: Response dynamics of wavemeter-locked Rydberg-state engagement and onset of dissipative time-crystal (DTC) oscillations.** (a) Spectrogram of the probe signal showing DTC oscillations as the coupler engages the $65S_{1/2}$ state from 0.43 GHz detuning. The fundamental and harmonic DTC oscillation bands appear shortly after each engagement (yellow dashed lines). (b) Corresponding wavemeter (WLM) frequency traces for the same sequence, where each step adjusts the PI-controller ratio $K_I/K_P$ = 3-30. Grey dashed lines mark the PI engagement, and yellow dashed lines indicate the delayed emergence of stable DTC oscillations. Increasing $K_I/K_P$ accelerates convergence to the target frequency, reducing rise time. (c) Overlay of the turn-on transients extracted from (b) for different gain ratios, showing rise times of 1.11-0.066 s. Higher $K_I/K_P$ values yield faster stabilization up to when mild ringing is observed before lock settles.

These results establish a compact, cavity-free platform capable of digitally reconfiguring Rydberg states with sub-MHz precision and multi-GHz s$^{-1}$ acquisition rates. The re-emergence of dissipative time-crystal oscillations after each controlled hop confirms rapid recovery of coherent many-body dynamics following active frequency switching. Together, the wavemeter-locked Rydberg system represents a scalable route toward reconfigurable, multi-band quantum electrometers and adaptive field sensors operating across dynamically selectable Rydberg states.

**Figure 5: Scaling of Rydberg-state acquisition rate with controller gain ratio.** Measured lock acquisition rate for the $65S_{1/2}$ state as a function of the PI controller gain ratio $K_I/K_P$, with an initial detuning of $\Delta f$ = 0.4283 GHz. The acquisition rate increases as 0.08 $(K_I/K_P)^{1.2}$, indicating power-law dependence between integral-to-proportional gain ratio and frequency-locking speed. Higher $K_I/K_P$ values enable faster convergence to the target Rydberg-state.

ACKNOWLEDGMENTS

The research was carried out at the Jet Propulsion Laboratory, California Institute of Technology, under a contract with the National Aeronautics and Space Administration (80NM0018D0004), through the Instrument Incubator Program's (IIP) Instrument Concept Development (Task Order 80NM0022F0020).

AUTHOR CONTRIBUTIONS

D.A. proposed the project. D.A. configured the atomic systems to include lasers, and stabilization/locking systems. D.A. developed the digital system for wavemeter locking and laser error reduction. D.A. developed and optimized excitation and field sources for DTC structure observation. D.A. designed the software scripts for data collection and processed the data for the figures. D.A. conducted all data collection efforts. D.A prepared the manuscript.

DATA AVAILABILITY STATEMENT

The frequency-domain slices of the readout at times: t = 10 s (blue) shows noise before DTC engagement; t = 40 s (yellow) and t = 55 s (red), given in Fig. 3c, is available as source data. The overlay of the turn-on transients extracted for different gain ratios (showing rise times of 1.11-0.066 s), given in Fig. 4c, is available as source data. All other data is available upon reasonable request to the corresponding author, Darmindra Arumugam via email: darmindra.d.arumugam@jpl.nasa.gov.

# Rydberg-State Hopping in a Wavemeter-Locked Dissipative Time-Crystal System